# Impact of Device Resistances in the Performance of Graphene-based Terahertz Photodetectors

O. Castelló[1,2] [0009-0009-8889-6270], Sofía M. López Baptista[1], K. Watanabe[3] [0000-0003-3701-8119], T. Taniguchi[4] [0000-0002-1467-3105], E. Diez[5] [0000-0001-7964-4148], J.E. Velázquez-Pérez[1,5] [0000-0002-6555-9871], Y.M. Meziani[1,5] [0000-0001-5193-7993], J.M. Caridad[1,2,a) [0000-0001-8943-1170] and J.A. Delgado-Notario[1,5 a) [0000-0001-9714-8180]

[1] *Department of Applied Physics, University of Salamanca, 37008 Salamanca, Spain*

[2] *Unidad de Excelencia en Luz y Materia Estructurada (LUMES), University of Salamanca, Spain*

[3] *Research Center for Electronic and Optical Materials, National Institute for Materials Science, 1-1 Namiki, Tsukuba 305-0044, Japan*

[4] *Research Center for Materials Nanoarchitectonics, National Institute for Materials Science, 1-1 Namiki, Tsukuba 305-0044, Japan*

[5] *Nanotechnology Group, USAL–Nanolab, Universidad de Salamanca, E-37008 Salamanca, Spain*

a) *Author to whom correspondence should be addressed: jose.caridad@usal.es and juanandn@usal.es*

**Abstract.** In recent years, graphene Field-Effect-Transistors (GFETs) have demonstrated an outstanding potential for Terahertz (THz) photodetection due to their fast response and high-sensitivity. Such features are essential to enable emerging THz applications, including 6G wireless communications, quantum information, bioimaging and security. However, the overall performance of these photodetectors may be utterly compromised by the impact of internal resistances presented in the device, so-called access or parasitic resistances. In this work, we provide a detailed study of the influence of internal device resistances in the photoresponse of high-mobility dual-gate GFET detectors. Such dual-gate architectures allow us to fine tune (decrease) the internal resistance of the device by an order of magnitude and consequently demonstrate an improved responsivity and noise-equivalent-power values of the photodetector, respectively. Our results can be well understood by a series resistance model, as shown by the excellent agreement found between the experimental data and theoretical calculations. These findings are therefore relevant to understand and improve the overall performance of existing high-mobility graphene photodetectors.

**Keywords:** Graphene, THz, Photodetector, Field-effect transistor, Plasmonic

# 1. INTRODUCTION

Electromagnetic radiation at Terahertz (THz) frequencies (0.1 – 10 THz) exhibit key advantageous features including a noninvasive nature (with energies of the order of few meVs) and penetration capability through opaque objects to the visible light. Such characteristics makes THz waves an amazing tool that can be used effectively in a wide variety of applications, e.g., upcoming high-speed wireless communication, medical bioimaging, information technologies, security, spectroscopy or even spintronics[1–5]. A general requirement for all of these technologies is the development of efficient photodetectors with a high sensitivity, low signal-to-noise ratio, and a fast response in the THz spectral range[6]. In recent years, the emergence of two-dimensional (2D) materials has boosted the development of novel prototype detectors and sensors operating at THz wavelengths due to their ability to exhibit striking, tunable optoelectronic properties. To date, a wide variety of 2D systems have been studied and exploited at THz wavelengths, including graphene[7, 8], Transition Metal Dichalcogenides (TMDs)[9], black phosphorus (BP)[10], topological semimetals[11] or even MXene[12]. The capacity of these photodetectors to convert incident THz photons into electrical signals was based on different physical mechanisms, including photo-thermoelectric, bolometric, or even photogalvanic effects[7, 8, 10]. However, plasma-wave assisted mechanisms in graphene field-effect transistors (FETs) detectors (also known as plasmonic graphene FETs or plasmonic graphene photodetectors) represent the most promising route to fulfill all the required criteria to be efficiently used in real applications[13].

Up to now, it has been extensively reported how the sensitivity of plasmonic THz detectors depends strongly on several factors including the integration of external components such as i) efficient plasmonic antennas to produce optical-field enhancement[14], ii) hyper-hemispherical lenses[15] and mesoscale particles[16] to improve the coupling of the THz beam and even focusing at subwavelength frequencies, iii) the presence of asymmetry in the device structure[17] and even iv) impedances presented in read-out circuits[18–20]. Similarly, the internal material's properties such as the mobility of free carriers in the channel[21] play a big role in the sensitivity of the device (reason why several FETs made from different materials have been studied so far[13, 22–25]). Nonetheless, the detail impact of additional (and relevant) intrinsic device parameters on the photogenerated signal at THz frequencies is less well-known in these systems.

In general, coupling of THz waves into plasmonic photodetectors based on FET architectures is effectively realized between a local (i.e. not covering the whole channel) top-gate and the source electrodes. The rectified DC current (or voltage) arising between drain and source electrodes when shining THz radiation to plasmonic photodetectors may be also heavily affected by parasitic regions in the channel (i.e. those regions in the channel not controlled by this local top-gate and not contributing in coupling the incoming THz radiation), and their impact on the overall photodetection performance of the system is not fully revealed yet. Such knowledge is important to completely understand and optimize the performance of FET devices at THz frequencies and this is the principal goal of the present study.

In order to do so, we have fabricated and analyzed the THz photoresponse of FET devices made from high-quality monolayer graphene, a very well-known semimetallic two-dimensional material with high-carrier mobility ($>50000$ cm$^2$V$^{-1}$s$^{-1}$) even when operating at room temperature. In practice, such high-quality samples are achieved by encapsulating the monolayer between hexagonal boron nitride (hBN)[26, 27]. The device FET architecture is similar to the one used in several studies in literature reporting plasmonic photodetection[25, 28–32], having a local top-gate (TG) electrode (which does not cover the entire device channel, see experimental details below). Moreover, the presence of a global back-gate electrode (BG) in the system allows us to independently dope channel regions not affected by the top-gate, tune the access resistance of these areas and therefore study their impact on the measured THz photoresponse.

In particular, by measuring the photocurrent collected in the device at different top- and back-gate potentials, we observe regimes with a notably enhanced photoresponse (more than one order of magnitude) when the areas of the channel not affected by the top-gate are highly doped. We further demonstrate how this behaviour can be fully understood by taking into account the regions with different doping levels existing in the photodetector in a series resistance model, showing an excellent agreement between this model and the experimental data.

## 2. RESULTS AND DISCUSSIONS

Figure 1 (a) shows the optical image of the studied graphene-FET (GFET) and a schematic lateral view of the channel structure. The GFET device consists on a monolayer graphene encapsulated between two thin hBN flakes stacked using a polymer-based method for the assembly of van der Waals heterostructures[24, 26]. The stack has been placed on top of a standard $SiO_2$/Si substrate, which serves as a global bottom-gate electrode. Device fabrication, including the definition of both channel and metal electrodes, has been undertaken by using standard electron-beam lithography, dry-etching and metal evaporation techniques[24, 26]. In addition, the GFET detector contains a bow-tie antenna connected to the source and top-gate electrode as labeled in Figure 1 (a) to efficiently coupling the THz radiation incident on the detector. In terms of geometry, the GFET device length is $L_{ch} = 6$ µm and average width of $W_{ch} \approx 6$ µm with a local top-gate covering only a part of the channel ($L_{TG} = 4.8$ µm). As such, there are two channel regions between drain-gate and source-gate electrodes not affected by the applied top-gate potentials. Importantly, the global bottom-gate electrode allows us to independently tune the doping level of these channel regions not covered by the top-gate and, therefore, to study their impact on the collected photocurrent in the detector. In other words, the access resistance of this device ($R_a$) is tunable in our device and is given by the contact resistance graphene-metal at the source and drain electrodes plus the resistance two regions not covered by the top-gate electrode (and therefore only affected by the back-gate).

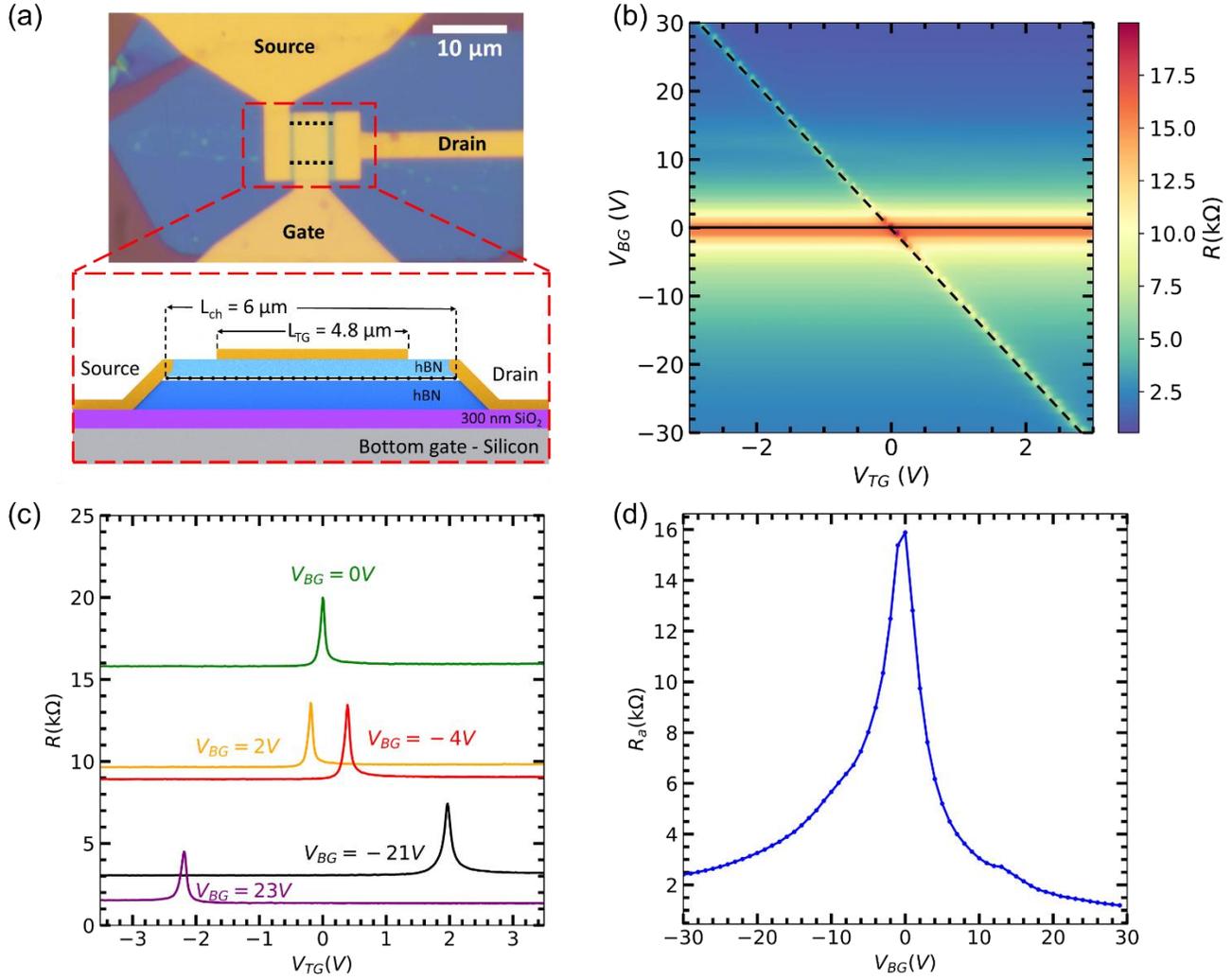

**Figure 1.** (a) Optical image (top) and lateral schematic view (bottom) of the dual gate Graphene FET. The black dotted line in the optical image indicates the edges of the graphene layer sandwiched between two h-BN flakes. (b) 2D map of the total channel resistance (R) as a function of the DC bias voltages applied to the top gate ($V_{TG}$) and to the bottom gate ($V_{BG}$) electrodes. The black horizontal (solid) and diagonal (dashed) lines correspond to the charge neutrality point (CNP) in non-top-gated and top-gated regions, respectively. (c) Total resistance as a function of the local top gate potential for five selected values of the back gate bias from panel (b). (d) Access resistance, $R_a$, as a function of the back gate bias. Temperature was fixed at 10K for panels (b)-(d).

Prior to study the photoresponse of our graphene photodetector, we have measured the electrical I-V characteristics of the device in a variable temperature optical cryostat included in the THz photocurrent setup (see Supplementary Material Note 1 for more information). Figure 1 (b) shows the device resistance (R) of the device, when sweeping simultaneously the top-gate voltage ($V_{TG}$) and back-gate voltage ($V_{BG}$) at a temperature of 10K. Meanwhile, Figure 1 (c) shows the total drain-to-source resistance of the device as a function of $V_{TG}$ for five selected back-gate potentials from Figure 1 (b). For simplicity and easier readability, the presented figures in this manuscript have been displayed as a function of the normalized back-gate voltage, i.e., $V_{BG} = V_{BG}^* - V_{BG,CNP}$, where $V_{BG}^*$ is the applied back-gate voltage and $V_{BG,CNP}$ is the back-gate voltage at the charge neutrality point (CNP). This means that in our device, the total device resistance reaches the maximum when both gates are biased such that the device is operating at the CNP potentials (i.e. $V_{BG} = 0V$ and $V_{TG} = 0V$), occurring at the crossing point of the continuous and dashed lines shown in Figure 1(b). When fixing the back-gate voltage at its CNP, the (electron or hole) carrier concentration increases in the channel region below the top-gated electrode when sweeping $V_{TG}$ resulting in the typical bell-shape resistance curve (see Figure 1 (c), green curve) characteristic of single-gate graphene devices[21, 30, 33]. A similar result occurs when $V_{BG}$ is biased at fixed potentials away from $V_{BG,CNP}$ (either positive or negative) and sweeping $V_{TG}$ (see Figure 1 (c)). However, one can see two clear main differences: First, the overall measured channel resistance curves are vertically shifted for $|V_{BG}| > 0V$ as a result of the modulation of the carrier concentration in the non-top-gate regions and, second, a shift of the $V_{TG,CNP}$ occurs as the region of the channel influenced by the top-gate electrode is also affected by the bottom gate electrode and therefore a new top-gate voltage must be applied to reach the new charge neutrality condition in this region[34].

As such, the total channel resistance, $R$, in the GFET can be understood as the sum of two different contributions[21], $R = R_{TG} + R_a$. The first term ($R_{TG}$) corresponds to the channel resistance below the top gated region (subjected to both local top- and bottom-gate potentials). The second contribution ($R_a$) stems from channel areas non influenced by the top-gate potential. As previously introduced, this second contribution, usually known as parasitic resistance or access resistance[21, 35], is given by $R_a = R_c + R_{nTG}$ and takes into account the sum of contact resistances ($R_c$) and the channel resistance of the non-top-gated areas ($R_{nTG}$). Both parasitic contributions limit the performance of any photodetector[21]. A good estimation of $R_a$ can be extracted from Figures 1 (b) and 1 (c) at large doping values (high $V_{TG}$, see Figure 1 (d)), i.e., when the total channel resistance R is mostly independent of $V_{TG}$ and has a constant value (i.e. $R \approx R_a$ for large top gate potentials, see Supplementary Material Note 2 for further details about how to extract $R_a$). Moreover, from the former transport data, we can also estimate the mobility of the channel at 10K with average mobilities exceeding 70.000 cm$^2$V$^{-1}$s$^{-1}$ (See Supplementary Material Note 2).

Next, we have analyzed the performance of the device when was exposed to radiation at a frequency of 0.3 THz. Figure 2 (a) shows the evolution of the measured photocurrent, $I_{PC}$, with respect to $V_{TG}$ and $V_{BG}$, and, Figure 2 (b) highlights $I_{PC}(V_{TG})$ for five selected back-gate potentials. For each fixed value of $V_{BG}$, $I_{PC}(V_{TG})$ exhibits an antisymmetric shape with respect to the applied top gate potential with a different sign depending on the type of carrier of the device channel as well as maximum ($I_{PC,max}$) and minimum ($I_{PC,min}$) values in the vicinity of the charge neutrality point $V_{TG,CNP}$ for hole and electron carriers, respectively. Moreover, the actual zero-crossing occurs right at the $V_{TG,CNP}$, and the photoresponse tends to zero at large $|V_{TG}|$ values. This qualitative behavior stems from the ambipolar charge transport presented in this material and is consistent with the Dyakonov-Shur (DS) detection mechanism (also referred to as resistive self-mixing mechanism). In this low-end THz frequency range, plasma waves injected from the source side may propagate along the channel of the FET at relatively high speed (plasma-wave propagation velocity in graphene ranges between $3 \cdot 10^6$ ms$^{-1}$ and $7 \cdot 10^6$ ms$^{-1}$ [36]). However, despite the large propagation speed, plasma oscillations are strongly overdamped and therefore the system responds in a quasi-static manner operating in the so-called broadband (non-resonant) regime in agreement with previous studies of THz photodetectors made of graphene GFETs [24, 30, 31] . The broadband operation of our photodetector is confirmed by noting that the value of the extracted relaxation time, $\tau$, of our graphene photodetector with an average mobility of 70.000 cm$^2$V$^{-1}$s$^{-1}$ at a carrier density of n $\approx 10^{15}$ m$^{-2}$ is approximately 0.3 ps. At 0.3 THz, this corresponds to a device quality factor below unity, Q = 0.57 (Q = $2\pi f\tau$, with f being the incoming THz frequency), typical of a THz photodetector working in the broadband regime [13, 23, 37], in a clear contrast with photodetectors operating at higher frequencies and exhibiting resonant detection (Q >> 1) at THz fields[29, 32].

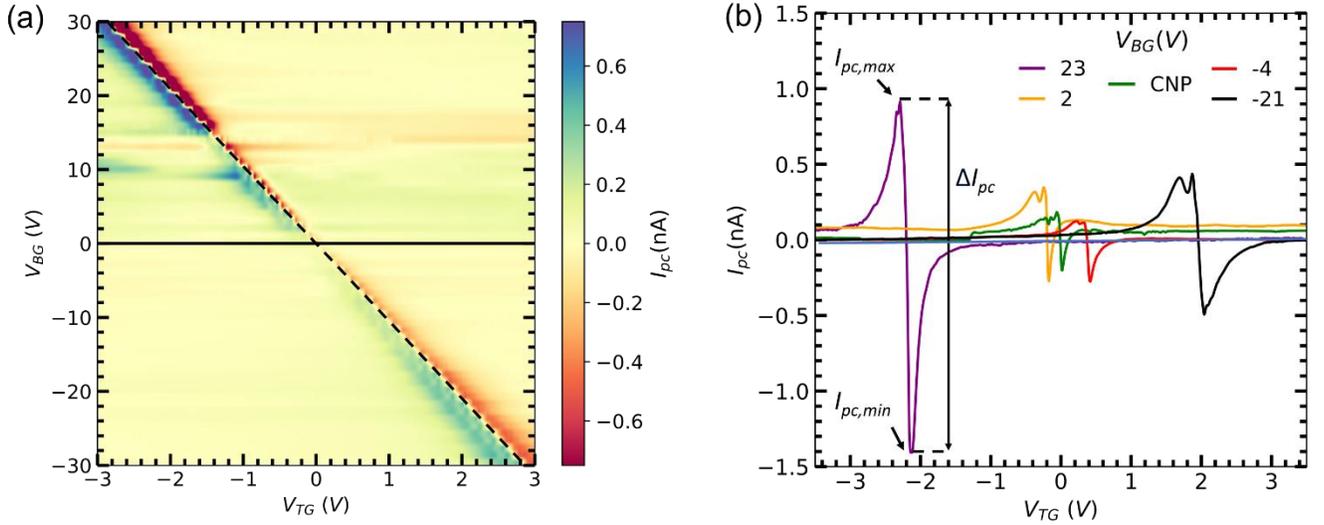

**Figure 2**. (a) Photocurrent ($I_{pc}$) mapping as function of top gate bias ($V_{TG}$) and back gate bias ($V_{BG}$) for an incident radiation of 0.3 THz at 10K. The black horizontal (solid) and diagonal (dashed) lines correspond to the CNP of the non-top-gated and top-gated regions, respectively. (b) $I_{pc}$ as a function of top gate voltage $V_{TG}$ for five selected values of a back gate bias from panel (a).

Interestingly, by examining Figure 2 in more quantitative terms, we found that the measured photocurrent $I_{PC}(V_{TG})$ has a strong dependence with the applied back-gate bias. As the back-gate potential is positioned away from the CNP ($V_{BG} = 0V$, solid horizontal line in Figure 2(a)), we observed a substantial increment in the value of both photocurrent maxima, $I_{PC,max}$, and minima, $I_{PC,min}$, and subsequently the absolute difference value of $|I_{PC,max} - I_{PC,min}|$ in the vicinity of the CNP of the top gate potential (diagonal dashed line in Figure 2(a)). In particular, we measure a maximum enhancement of $|I_{PC,max} - I_{PC,min}|$ at $V_{BG} >$ 20V which is 20 times larger than the value at $V_{BG} = 0$V. This fact indicates that the doping level of areas not covered by the top-gate potential notably influences the photodetection and therefore the overall performance of the graphene FET device, with larger photoresponses occurring for higher doping levels.

As shown below, the observed enhancement of the THz photocurrent can be interpreted and well modeled by the formerly introduced series resistance model $R_a = R_c + R_{nTG}$. From a theoretical perspective, the predicted DS photocurrent, $I_{pred}$, expected in plasmonic photodetectors when an oscillating THz field is coupled between gate and source electrodes and the device exhibit a broadband response, is given by the following expression[13, 28, 30]:

$$I_{pred} = -\frac{U_a^2}{4}\frac{d\sigma(V_G)}{dV_G}$$  (1)

where $U_a$ is the amplitude of the THz-induced ac voltage between the gate and the source electrodes, $\sigma$ is the experimental DC total conductance of the device ($\sigma = L_{ch}/W_{ch}/R$) and $V_G$ the corresponding potential at the gate electrode where the THz radiation is coupled onto the device ($V_{TG}$ in our device). Importantly, although commonly neglected, Eq. 1 does depend on the access resistance of the actual devices, playing an important role in the overall device performance. In more detail, since $R = R_{TG} + R_a$, the total channel conductance of the device, $\sigma$, can be expressed as:

$$\sigma = \frac{L_{ch}}{W_{ch}}\frac{\sigma_{TG}}{1 + R_a\sigma_{TG}}$$  (2)

where $\sigma_{TG}$ represent the channel conductance at the dual-gate region ($\sigma_{TG} = \frac{1}{R_{TG}}$, i.e., below the top-gate electrode)). In order to interpret precisely the observed experimental behavior of the photocurrent device (i.e. the fact that the maxima and minima values of $I_{PC}(V_{TG})$ decrease when $V_{BG}$ tends to $V_{BG,CNP}$ in Figure 2), the access resistance contribution must be introduced in the predicted photocurrent. Then, replacing Equation 2 into Equation 1, the predicted photocurrent when measuring as a function of the local top gate can be rewritten as:

$$I_{pred} = -\frac{U_a^2}{4}\frac{L_{ch}}{W_{ch}}\frac{d\sigma_{TG}}{dV_{TG}}\frac{1}{(1+R_a\sigma_{TG})^2} \tag{3}$$

From Equation (3), the expected photocurrent depends on both the conductance of the graphene channel below the top-gated region as well as on the access resistance. Moreover, it can be seen that the maximum current photoresponse as a function of the local top gate potential occurs when $R_a$ is reduced (i.e. contact and non-top-gate region resistances are minimized). Such reduced $R_a$ occurs for large values of $V_{BG}$, in clear qualitative agreement with our experimental data (Figure 2). Conversely, $R_a$ reaches its maximum value when the back-gate bias is set at the CNP ($V_{BG} = 0V$ in this letter), yielding to minimum photocurrent values at that gate-voltages (also in agreement with experiments in Figure 2).

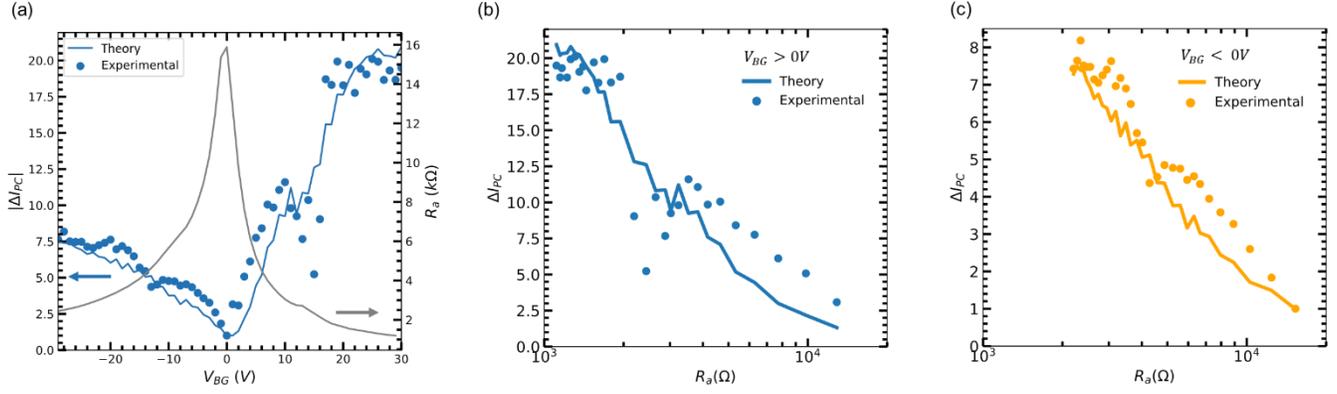

**Figure 3.** (a) Normalized $\Delta I_{pc} = I_{pc,max} - I_{pc,min}$ values (left axis) and access resistance (right axis) as a function of the back-gate voltage extracted from the measured photocurrent data (blue dots) and expected photocurrent using Eq. 3 (blue line). The normalization is carried out with respect to $\Delta I_{pc}$ at the back-gate $V_{BG,CNP}$. Normalized $\Delta I_{pc}$ values as a function of the access resistance for positive (b) and negative (c) back-gate potentials. Solid line represents the theoretical dependence with respect to $R_a$ using the Eq. 3 and dots the experimental values extracted from the measured photocurrent data. Temperature was fixed at 10K and frequency was 0.3 THz in all panels.

We further demonstrate the validity of Equation 3 in quantitative terms by comparing the observed photocurrent dependence as a function of the applied back- and top-gate voltages (Figure 2) with the predicted one using Equation 3 (see supplementary Material Note 3). Furthermore, we presented the difference between the maxima and minima of the photocurrent, $\Delta I_{pc} = |I_{pc,max} - I_{pc,min}|$ for the different applied $V_{BG}$ in both cases, the experimental photocurrent and the predicted photocurrent, using Eq. 3 (see Figure 3(a)) and analyzed its dependence with the modulation of the access resistance. The difference of maxima and minima of the measured photocurrent (blue dots in Figure 3 (a)) is in excellent agreement (qualitatively and quantitatively) with the data obtained when using Eq. 3 and the DC measured values in Figure 1 (blue line in Figure 3(a)). To perform this comparison, for each value of $V_{BG}$, $\Delta I_{pc}$ has been normalized with respect to the value obtained at the CNP (i.e. $V_{BG} = 0V$) and therefore Figure 3 represents the photocurrent enhancement ratio with respect to this operation point. Importantly, the maximum enhancement ratio exhibited in the system is as high as 20 when the access resistance is decreased by nearly one order of magnitude.

All these findings are summarized in Figures 3 (b) and 3(c), showing the enhancement of the measured photocurrent, $\Delta I_{pc}$, as a function of $R_a$ when doping the channel with both electron (positive values of $V_{BG}$) and hole (negative values of $V_{BG}$) carriers. The larger photoresponse (i.e the larger $|I_{pc,max} - I_{pc,min}|$) is obtained for the lower values of $R_a$ and the experimental trend is in clear agreement with the expected photoresponse using Eq. 3 (solid lines in Figure 3 (b) and (c)). In fact, we showed that the rectified photocurrent in plasmonic photodetectors at THz frequencies depends on the access resistance as $\sim a/(1 + bR_a)^2$, where $a$ and $b$ are parameters that are independent of $R_a$. The above presented results are essential findings for the optimal performance and guidance in the design of any state-of-the-art plasmonic THz photodetectors.

Prior to conclude, two figures of merits define the sensitivity of a photodetector, i) the current responsivity ($R_I = I_{pc}/P_d$, where $P_d$ is the incoming power at the detector) and ii) the noise-equivalent-power ($NEP = \sqrt{4k_B T\sigma}/R_I$ for a non-drain-to-source-biased detector where $k_B$ is the Boltzmann constant, and $T$ is the temperature). Then high-responsivity and low NEP values are desirable in any THz photodetector. In terms of the responsivity, it is clear that the higher measured values of $I_{pc}$, the better current responsivity. Thus, as we show above, an optimal performance device (i.e. the modulation of the access resistance through the back-gate bias) can increase by 20 times the rectified THz photocurrent, resulting in a direct enhancement of the responsivity by the same factor. Similarly, the modulation of the access resistance towards lower values yields to lower values of NEP (see Supplementary Material Note 4).

From application perspective, although the aforementioned analysis is shown at a cryogenic temperature (10K), a similar behaviour occurs at application-relevant, room temperature conditions where the above introduced relevant figure of merits (Responsivity and NEP) can be also enhanced by modulating and minimizing the access resistance (See Supplementary Material Note 5). While we have shown that access resistance can be easily modulated by using a dual-gate configuration, if the substrate characteristics hinder the implementation of an additional global-back-gate[38], the access resistance contribution may be minimized by reducing the gap between drain-gate and source-gate electrodes or even setting in these regions a p-type or n-type doping in graphene achieved through chemical doping[39] which will be beneficial in terms of the THz photodetector performance.

Finally, is important to remark that the performance of THz detectors depends not only on internal resistances but also on the characteristics of the photodetector's active materials. For this reason, our study has been focused on single-layer graphene, which has been extensively studied in recent years due to its superior carrier mobility, low resistance, gapless spectrum, and ultrafast carrier dynamics. These outstanding optoelectronic properties result in the fastest response times with high responsivity at THz frequencies[40]. However, graphene is just one of many possible 2D crystals, and our findings can be extended to improve and study the performance of sensors made from other 2D materials.

For instance, among the different 2D materials, BP can support relatively high carrier mobility at room temperature (around $10^3 \, \mathrm{cm^2 V^{-1} s^{-1}}$) making it suitable for the development of high-frequency plasmonic THz detectors[10, 41]. However, in contrast to graphene, BP degrades rapidly in air, affecting its structure and properties, and ultimately constrain the reproducibility of any study. Therefore, significant effort is still required to exploit its potential use in high-frequency applications.

Alternative layered 2D materials such as TMDs have also been studied in THz science due to their high modulation efficiency[9, 41] compared to graphene. However, their relatively poor mobility remains a major constraint for high-frequency plasmonic photodetectors. Additionally, their wide intrinsic band gaps (in the order of a few eVs) render them unsuitable for THz applications, which is why most studies are focused on Near Infrared (NIR) and visible frequencies[40].

Furthermore, topological semimetals have not been commonly considered for photodetection due to their tendency to suffer from high dark current. However, they are intriguing materials, with carrier mobility values that, in some cases, could overcome those of graphene[41]. These novel materials present a wide range of physical phenomena that could pave the way for their applications in optoelectronics and novel photodetectors at THz wavelengths[11].

## 3. CONCLUSIONS

In summary, we have observed experimentally that the THz photoresponse in graphene FET photodetectors is notably affected by the access resistance, including metal-graphene resistances and channel areas not affected by local gate electrodes. In particular, the measured rectified THz photoresponse increases one order of magnitude in the photodetector when minimizing $R_a$, which highlights the importance of reducing any internal resistance in the device. All these findings are well captured in qualitative and quantitative terms can be captured by a series resistance model of the device. In this context, our findings, together with the optimization of i) the plasmonic antenna and external optical elements to efficiently concentrate the electromagnetic energy into the graphene channel[14], ii) low contact resistances and iii) ultra-clean van der Waals samples with high-mobility[27], would set new strategies for the development of state-of-the-art THz detectors operating at room-temperature with high optical responsivity, low values of NEP and fast response.

## 4. SUPPLEMENTARY MATERIAL

See Supplementary Material for additional details of the device mobility, photoresponse experiments at different temperatures and measurements of the device noise equivalent power.

## 5. ACKNOWLEDGMENTS

Authors thank the support from the Ministry of Science and Innovation (MCIN) and the Spanish State Research Agency (AEI) under grants (PID2021-126483OB-I00, PID2021-128154NA-I00 and PID2022-136285NB-C32) funded by MCINU/AEI/10.13039/501100011033 and by "ERDF A way of making Europe". This work has been also supported by Junta de Castilla y León co-funded by FEDER under the Research Grant numbers SA103P23 and SA106P23. K.W. and T.T. acknowledge support from the JSPS KAKENHI (Grant Numbers 21H05233 and 23H02052) and World Premier International Research Center Initiative (WPI), MEXT, Japan. J.M.C acknowledges financial support by the MCIN and AEI "Ramón y Cajal" program (RYC2019-028443-I) funded by MCINU/AEI/10.13039/501100011033 and by "ESF Investing in Your Future". J.M.C also acknowledges financial of the European Research Council (ERC) under Starting grant ID 101039754, CHIROTRONICS, funded by the European Union. Views and opinions expressed are however those of the author(s) only and do not necessarily reflect those of the European Union or the European Research Council. Neither the European Union nor the granting authority can be held responsible for them. J.A.D-N thanks the support from the Universidad de Salamanca for the María Zambrano postdoctoral grant funded by the Next Generation EU Funding for the Requalification of the Spanish University System 2021–23, Spanish Ministry of Universities. Authors also acknowledge USAL-NANOLAB for the use of Clean Room facilities.

## 6. REFERENCES

1.  Heffernan BM, Kawamoto Y, Maekawa K, Greenberg J, Amin R, Hori T, Tanigawa T, Nagatsuma T, Rolland A (2023) 60 Gbps real-time wireless communications at 300 GHz carrier using a Kerr microcomb-based source. APL Photonics 8:. https://doi.org/10.1063/5.0146957/18004712/066106_1_5.0146957.PDF

2.  Yu L, Hao L, Meiqiong T, Jiaoqi H, Wei L, Jinying D, Xueping C, Weiling F, Yang Z (2019) The medical application of terahertz technology in non-invasive detection of cells and tissues: opportunities and challenges. RSC Adv 9:9354–9363. https://doi.org/10.1039/C8RA10605C

3.  Wang Y, Li W, Cheng H, Liu Z, Cui Z, Huang J, Xiong B, Yang J, Huang H, Wang J, Fu Z, Huang Q, Lu Y (2023) Enhancement of spintronic terahertz emission enabled by increasing Hall angle and interfacial skew scattering. Communications Physics 2023 6:1 6:1–10. https://doi.org/10.1038/s42005-023-01402-x

4.  Aghoutane B, Ghzaoui M El, Kumari S V., Das S, Faylali H El (2023) A circularly polarized super wideband transparent optical nanoantenna for advanced THz communication applications. Opt Quantum Electron 55:1–17. https://doi.org/10.1007/S11082-022-04484-Z/TABLES/2

5.  Sun Z, Liang C, Chen C, Wang X, Zhou E, Bian X, Yang Y, You R, Zhao X, Zhao J, You Z (2024) High-Efficiency Dynamic Terahertz Deflector Utilizing a Mechanically Tunable Metasurface. Research 6:0274. https://doi.org/10.34133/research.0274

6.  Leitenstorfer A, Moskalenko AS, Kampfrath T, Kono J, Castro-Camus E, Peng K, Qureshi N, Turchinovich D, Tanaka K, Markelz AG, Havenith M, Hough C, Joyce HJ, Padilla WJ, Zhou B, Kim KY, Zhang XC, Jepsen PU, Dhillon S, Vitiello M, Linfield E, Davies AG, Hoffmann MC, Lewis R, Tonouchi M, Klarskov P, Seifert TS, Gerasimenko YA, Mihailovic D, Huber R, Boland JL, Mitrofanov O, Dean P, Ellison BN, Huggard PG, Rea SP, Walker C, Leisawitz DT, Gao JR, Li C, Chen Q, Valušis G, Wallace VP, Pickwell-MacPherson E, Shang X, Hesler J, Ridler N, Renaud

CC, Kallfass I, Nagatsuma T, Zeitler JA, Arnone D, Johnston MB, Cunningham J (2023) The 2023 terahertz science and technology roadmap. J Phys D Appl Phys 56:223001. https://doi.org/10.1088/1361-6463/ACBE4C

7.  Otteneder M, Hubmann S, Lu X, Kozlov DA, Golub LE, Watanabe K, Taniguchi T, Efetov DK, Ganichev SD (2020) Terahertz Photogalvanics in Twisted Bilayer Graphene Close to the Second Magic Angle. Nano Lett 20:7152–7158. https://doi.org/10.1021/acs.nanolett.0c02474

8.  Castilla S, Terrés B, Autore M, Viti L, Li J, Nikitin AY, Vangelidis I, Watanabe K, Taniguchi T, Lidorikis E, Vitiello MS, Hillenbrand R, Tielrooij K-J, Koppens FHL (2019) Fast and Sensitive Terahertz Detection Using an Antenna-Integrated Graphene pn Junction. Nano Lett 19:2765–2773. https://doi.org/10.1021/acs.nanolett.8b04171

9.  Xie Y, Liang F, Chi S, Wang D, Zhong K, Yu H, Zhang H, Chen Y, Wang J (2020) Defect Engineering of MoS2 for Room-Temperature Terahertz Photodetection. ACS Appl Mater Interfaces 12:7351–7357. https://doi.org/10.1021/acsami.9b21671

10. Viti L, Hu J, Coquillat D, Politano A, Knap W, Vitiello MS (2016) Efficient Terahertz detection in black-phosphorus nano-transistors with selective and controllable plasma-wave, bolometric and thermoelectric response. Sci Rep 6:20474. https://doi.org/10.1038/srep20474

11. Hu Z, Zhang L, Chakraborty A, D'Olimpio G, Fujii J, Ge A, Zhou Y, Liu C, Agarwal A, Vobornik I, Farias D, Kuo C-N, Lue CS, Politano A, Wang S-W, Hu W, Chen X, Lu W, Wang L (2023) Terahertz Nonlinear Hall Rectifiers Based on Spin-Polarized Topological Electronic States in 1T-CoTe2. Advanced Materials 35:2209557. https://doi.org/https://doi.org/10.1002/adma.202209557

12. Yang S, Lin Z, Wang X, Huang J, Yang R, Chen Z, Jia Y, Zeng Z, Cao Z, Zhu H, Hu Y, Li E, Chen H, Wang T, Deng S, Gui X (2024) Stretchable, Transparent, and Ultra-Broadband Terahertz Shielding Thin Films Based on Wrinkled MXene Architectures. Nanomicro Lett 16:165. https://doi.org/10.1007/s40820-024-01365-w

13. Dyakonov M, Shur M (1996) Detection, mixing, and frequency multiplication of terahertz radiation by two-dimensional electronic fluid. IEEE Trans Electron Devices 43:380–387. https://doi.org/10.1109/16.485650

14. Wang L, Han L, Guo W, Zhang L, Yao C, Chen Z, Chen Y, Guo C, Zhang K, Kuo C-N, Lue CS, Politano A, Xing H, Jiang M, Yu X, Chen X, Lu W (2022) Hybrid Dirac semimetal-based photodetector with efficient low-energy photon harvesting. Light Sci Appl 11:53. https://doi.org/10.1038/s41377-022-00741-8

15. Yao C, Jiang M, Wang D, Zhang L, Zhang N, Wang L, Chen X (2023) Hemispherical lens integrated room temperature ultra-broadband GaAs HEMT terahertz detector. Front Phys 11:1182059. https://doi.org/10.3389/FPHY.2023.1182059/BIBTEX

16. Nguyen Pham HH, Hisatake S, Minin OV, Nagatsuma T, Minin IV (2017) Enhancement of spatial resolution of terahertz imaging systems based on terajet generation by dielectric cube. APL Photonics 2:. https://doi.org/10.1063/1.4983114/122982

17. Shabanov A, Moskotin M, Belosevich V, Matyushkin Y, Rybin M, Fedorov G, Svintsov D (2021) Optimal asymmetry of transistor-based terahertz detectors. Appl Phys Lett 119:163505. https://doi.org/10.1063/5.0063870/40444

18. Sakowicz M, Lifshits MB, Klimenko OA, Schuster F, Coquillat D, Teppe F, Knap W (2011) Terahertz responsivity of field effect transistors versus their static channel conductivity and loading effects. J Appl Phys 110:54512. https://doi.org/10.1063/1.3632058/991776

19. Stillman W, Shur MS, Veksler D, Rumyantsev S, Guarin F (2007) Device loading effects on nonresonant detection of terahertz radiation by silicon MOSFETs. Electron Lett 43:422–423. https://doi.org/10.1049/EL:20073475

20. Stillman W, Donais C, Rumyantsev S, Shur M, Veksler D, Hobbs C, Smith C, Bersuker G, Taylor W, Jammy R (2012) SILICON FINFETS AS DETECTORS OF TERAHERTZ AND SUB-

TERAHERTZ RADIATION. https://doi.org/101142/S0129156411006374 20:27–42. https://doi.org/10.1142/S0129156411006374

21. Rehman A, Delgado-Notario JA, Sai P, But DB, Prystawko P, Ivonyak Y, Cywinski G, Knap W, Rumyantsev S (2022) Temperature dependence of current response to sub-terahertz radiation of AlGaN/GaN and graphene transistors. Appl Phys Lett 121:213503. https://doi.org/10.1063/5.0129507

22. Hou HW, Liu Z, Teng JH, Palacios T, Chua SJ (2017) High Temperature Terahertz Detectors Realized by a GaN High Electron Mobility Transistor. Scientific Reports 2017 7:1 7:1–6. https://doi.org/10.1038/srep46664

23. Knap W, Dyakonov M, Coquillat D, Teppe F, Dyakonova N, Łusakowski J, Karpierz K, Sakowicz M, Valusis G, Seliuta D, Kasalynas I, El Fatimy A, Meziani YM, Otsuji T (2009) Field Effect Transistors for Terahertz Detection: Physics and First Imaging Applications. J Infrared Millim Terahertz Waves 30:1319–1337. https://doi.org/10.1007/s10762-009-9564-9

24. Delgado-Notario JA, Knap W, Clericò V, Salvador-Sánchez J, Calvo-Gallego J, Taniguchi T, Watanabe K, Otsuji T, Popov V V, Fateev D V, Diez E, Velázquez-Pérez JE, Meziani YM (2022) Enhanced terahertz detection of multigate graphene nanostructures. Nanophotonics 11:519–529. https://doi.org/doi:10.1515/nanoph-2021-0573

25. Viti L, Coquillat D, Politano A, Kokh KA, Aliev ZS, Babanly MB, Tereshchenko OE, Knap W, Chulkov E V, Vitiello MS (2016) Plasma-Wave Terahertz Detection Mediated by Topological Insulators Surface States. Nano Lett 16:80–87. https://doi.org/10.1021/acs.nanolett.5b02901

26. Delgado-Notario JA, Clericò V, Diez E, Velázquez-Pérez JE, Taniguchi T, Watanabe K, Otsuji T, Meziani YM (2020) Asymmetric dual-grating gates graphene FET for detection of terahertz radiations. APL Photonics 5:066102. https://doi.org/10.1063/5.0007249

27.	Vaquero D, Clericò V, Schmitz M, Delgado-Notario JA, Martín-Ramos A, Salvador-Sánchez J, Müller CSA, Rubi K, Watanabe K, Taniguchi T, Beschoten B, Stampfer C, Diez E, Katsnelson MI, Zeitler U, Wiedmann S, Pezzini S (2023) Phonon-mediated room-temperature quantum Hall transport in graphene. Nat Commun 14:318. https://doi.org/10.1038/s41467-023-35986-3

28.	Bandurin DA, Gayduchenko I, Cao Y, Moskotin M, Principi A, Grigorieva I V, Goltsman G, Fedorov G, Svintsov D (2018) Dual origin of room temperature sub-terahertz photoresponse in graphene field effect transistors. Appl Phys Lett 112:141101. https://doi.org/10.1063/1.5018151

29.	Bandurin DA, Svintsov D, Gayduchenko I, Xu SG, Principi A, Moskotin M, Tretyakov I, Yagodkin D, Zhukov S, Taniguchi T, Watanabe K, Grigorieva I V, Polini M, Goltsman GN, Geim AK, Fedorov G (2018) Resonant terahertz detection using graphene plasmons. Nat Commun 9:5392. https://doi.org/10.1038/s41467-018-07848-w

30.	Vicarelli L, Vitiello MS, Coquillat D, Lombardo A, Ferrari AC, Knap W, Polini M, Pellegrini V, Tredicucci A (2012) Graphene field-effect transistors as room-temperature terahertz detectors. Nat Mater 11:865–871. https://doi.org/10.1038/nmat3417

31.	Zak A, Andersson MA, Bauer M, Matukas J, Lisauskas A, Roskos HG, Stake J (2014) Antenna-Integrated 0.6 THz FET Direct Detectors Based on CVD Graphene. Nano Lett 14:5834–5838. https://doi.org/10.1021/nl5027309

32.	Caridad JM, Castelló Ó, López Baptista SM, Taniguchi T, Watanabe K, Roskos HG, Delgado-Notario JA (2024) Room-Temperature Plasmon-Assisted Resonant THz Detection in Single-Layer Graphene Transistors. Nano Lett 24:935–942. https://doi.org/10.1021/acs.nanolett.3c04300

33.	Wang L, Meric I, Huang PY, Gao Q, Gao Y, Tran H, Taniguchi T, Watanabe K, Campos LM, Muller DA, Guo J, Kim P, Hone J, Shepard KL, Dean CR (2013) One-Dimensional Electrical Contact to a Two-Dimensional Material. Science (1979) 342:614–617. https://doi.org/10.1126/science.1244358

34.     Kim S, Nah J, Jo I, Shahrjerdi D, Colombo L, Yao Z, Tutuc E, Banerjee SK (2009) Realization of a high mobility dual-gated graphene field-effect transistor with Al2O3 dielectric. Appl Phys Lett 94:062107. https://doi.org/10.1063/1.3077021

35.     Boubanga-Tombet S, Knap W, Yadav D, Satou A, But DB, Popov V V, Gorbenko I V, Kachorovskii V, Otsuji T (2020) Room-Temperature Amplification of Terahertz Radiation by Grating-Gate Graphene Structures. Phys Rev X 10:31004. https://doi.org/10.1103/PhysRevX.10.031004

36.     Soltani A, Kuschewski F, Bonmann M, Generalov A, Vorobiev A, Ludwig F, Wiecha MM, Čibiraitė D, Walla F, Winnerl S, Kehr SC, Eng LM, Stake J, Roskos HG (2020) Direct nanoscopic observation of plasma waves in the channel of a graphene field-effect transistor. Light: Science & Applications 2020 9:1 9:1–7. https://doi.org/10.1038/s41377-020-0321-0

37.     Muraviev A V., Rumyantsev SL, Liu G, Balandin AA, Knap W, Shur MS (2013) Plasmonic and bolometric terahertz detection by graphene field-effect transistor. Appl Phys Lett 103:181114. https://doi.org/10.1063/1.4826139/130051

38.     Tamura K, Tang C, Ogiura D, Suwa K, Fukidome H, Takida Y, Minamide H, Suemitsu T, Otsuji T, Satou A (2022) Fast and sensitive terahertz detection with a current-driven epitaxial-graphene asymmetric dual-grating-gate field-effect transistor structure. APL Photonics 7:126101. https://doi.org/10.1063/5.0122305/2835249

39.     Ullah S, Shi Q, Zhou J, Yang X, Ta HQ, Hasan M, Mahmood Ahmad N, Fu L, Bachmatiuk A, Rümmeli MH, Ullah S, Shi Q, Zhou J, Yang X, Bachmatiuk A, Rümmeli MH, Ta HQ, Hasan M, Ahmad NM, Fu L (2020) Advances and Trends in Chemically Doped Graphene. Adv Mater Interfaces 7:2000999. https://doi.org/10.1002/ADMI.202000999

40.     Rogalski A (2019) Graphene-based materials in the infrared and terahertz detector families: a tutorial. Adv Opt Photonics 11:314–379. https://doi.org/10.1364/AOP.11.000314

41.   Yang J, Qin H, Zhang K (2018) Emerging terahertz photodetectors based on two-dimensional materials. Opt Commun 406:36–43. https://doi.org/https://doi.org/10.1016/j.optcom.2017.05.041

**Supplementary Material: Impact of Device Resistances in the Performance of Graphene-based Terahertz Photodetectors**

O. Castelló[1,2], Sofía M. López Baptista[1], K. Watanabe[3], T. Taniguchi[4], E. Diez[5], J.E. Velázquez-Pérez[1,5], Y.M. Meziani[1,5], J.M. Caridad[1,2,a)] and J.A. Delgado-Notario[1,5 a)]

[1] *Department of Applied Physics, University of Salamanca, 37008 Salamanca, Spain*

[2] *Unidad de Excelencia en Luz y Materia Estructurada (LUMES), University of Salamanca, Spain*

[3] *Research Center for Electronic and Optical Materials, National Institute for Materials Science, 1-1 Namiki, Tsukuba 305-0044, Japan*

[4] *Research Center for Materials Nanoarchitectonics, National Institute for Materials Science, 1-1 Namiki, Tsukuba 305-0044, Japan*

[5] *Nanotechnology Group, USAL–Nanolab, Universidad de Salamanca, E-37008 Salamanca, Spain*

[a)] *Author to whom correspondence should be addressed:* jose.caridad@usal.es *and* juanandn@usal.es

**Supplementary Material Note 1: Measurement set-ups**

DC transport measurements were performed using a Standford SR860 lock-in amplifier with an excitation quasi-DC current of 10nA at a frequency of 11.3 Hz. The generation of the THz wave is carried out by an RPG dual frequency THz source. This source generates a signal of 0.3 THz and 6 mW of output power using a solid-state dielectric resonator and amplifier stages based on Schottky diodes. This signal is modulated at 333 Hz mechanical chopper and them is collimated and focused on the device using an optical system. The dual gate GFET is inside a cryostat that allows us to maintain the device at a temperature between 4.5K and 300K. The window of the cryostat is made from teflon, which is transparent to THz radiation. The THz signal generates a DC current in the transistor via a nonlinear mechanism. The photocurrent was measured using a low DC input impedance current to voltage preamplifier SR570 and a lock-in amplifier SR860 whose reference follows the frequency of the chopper. Also, to apply a top and back gate bias and measure the leakage current through the gate we use a double channel Keithley 2412 source-meter. For more information, the reader is referred to previous studies[1].

**Supplementary Material Note 2: Mobility estimation.**

From experimental DC measurements between the gate and drain contacts as function of $V_{TG}$ we can extract the mean mobility ($\mu$) and account the resistance not affected by the top-gate ($R_2$) of electrons and holes. The total resistance of the device (drain to source) can be calculated from[2].

$$R = R_a + \frac{N_{sq}}{n_{tot} e \mu} \tag{S1}$$

where $R_a$ is the access resistance, that corresponds to the sum of the resistance in the regions without top gate and the contact resistances (drain and source electrodes), $N_{sq}$ is the number of squares of the graphene channel, $\mu$ is the mean mobility of the carriers in the graphene channel, and $n_{tot}$ is the total density of carriers that can be calculated with:

$$n_{tot} = \sqrt{n^2 + n_0^2} \tag{S2}$$

where $n$ is the charge density induced by the top gate and $n_0$ the residual charge carrier density:

$$n = \frac{c_T}{e}\left(V_{TG} - V_{TG}^{(CNP)}\right) \tag{S3}$$

To reduce the problem to a two-parameter adjustment, it has been used that[3]:

$$n_0 = \frac{N_{sq}}{\left(R^{(CNP)} - R_a\right) e \mu} \tag{S4}$$

Former expression that can be obtained from equation S1 assuming that $n_0$ correspond to the charge density at the Dirac point, where $R = R^{(CNP)}$. Combining eq S1, S2 and eq S4 we can obtain the equation described by Gammelgaard et al[3]:

$$R = R_a + \frac{N_{sq}}{\sqrt{n^2 + \left(\frac{N_{sq}}{\left(R^{(CNP)} - R_a\right)e\mu}\right)^2} \, e\mu} \tag{S5}$$

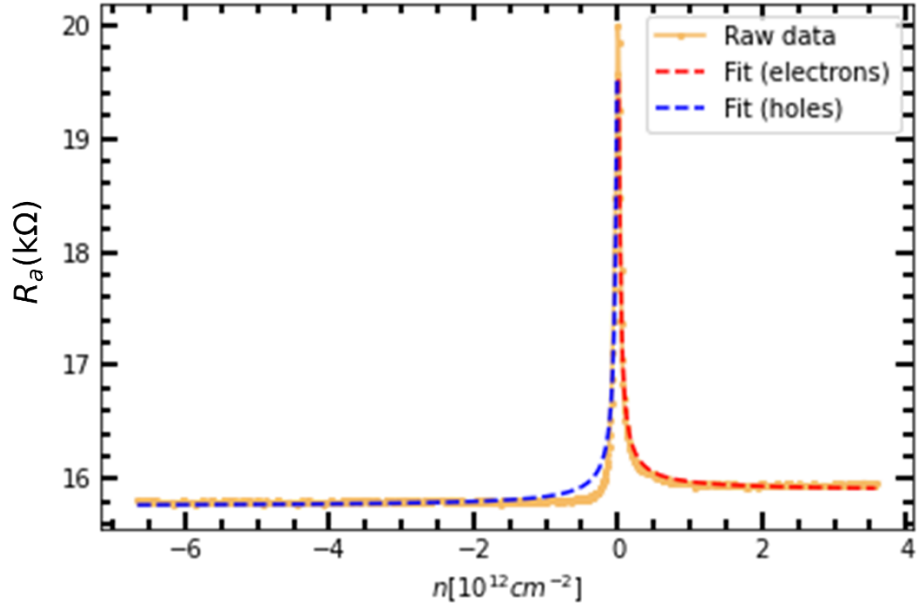

**Figure S1.** Experimental curve fitting of the resistance (orange line) versus carrier density for a $\boldsymbol{V_{BG}} = \boldsymbol{0V}$. The red dashed line corresponds to the fitting made for the electron branch and the blue dashed line for holes.

The extracted values at 10K temperature for mobilities, $\mu$, were around 76000 cm$^2$V$^{-1}$s$^{-1}$ and 71000 cm$^2$V$^{-1}$s$^{-1}$ for electrons and holes respectively. The extracted $R_a$ for electrons (holes) was 15893 $\Omega$ (15756 $\Omega$).

Additionally, as mentioned in the main text, we estimated the value of the access resistance $(R_a)$ by recording the value of $R$ at large top-gate potentials. Furthermore, $R_a$ is also estimated by fitting the measured $R(V_{TG})$ curve to an expression with two free parameters $R_a$ and μ (see equation S5). The discrepancy between both methods is rather small, below 15% (see Table S1 below) which confirm the estimation of $R_a$ as the value of $R$ at large $V_{TG}$.

| | Value away from V$_{CNP}$ | | Fit to model | |
|---|---|---|---|---|
| $V_{BG}$ (V) | $R_a$ (Ω) electrons) | $R_a$ (Ω) (holes) | $R_a$ (Ω) electrons) | $R_a$ (Ω) (holes) |
| 17 | 1941 | 2349 | 1937 | 2266 |
| 30 | 1113 | 1272 | 1071 | 1203 |
| 0 | 15463 | 15297 | 15408 | 15233 |
| -28 | 2557 | 2343 | 2251 | 2245 |

**Table S1.** Data comparison between possible access resistance extraction methods. On the left are the values corresponding to the value when we move far enough from the Dirac point (i.e. large top gate voltages) and on the right are those extracted through the adjustment procedure following equation S5 described above in this section.

## Supplementary Material Note 3: Theoretical THz photocurrent.

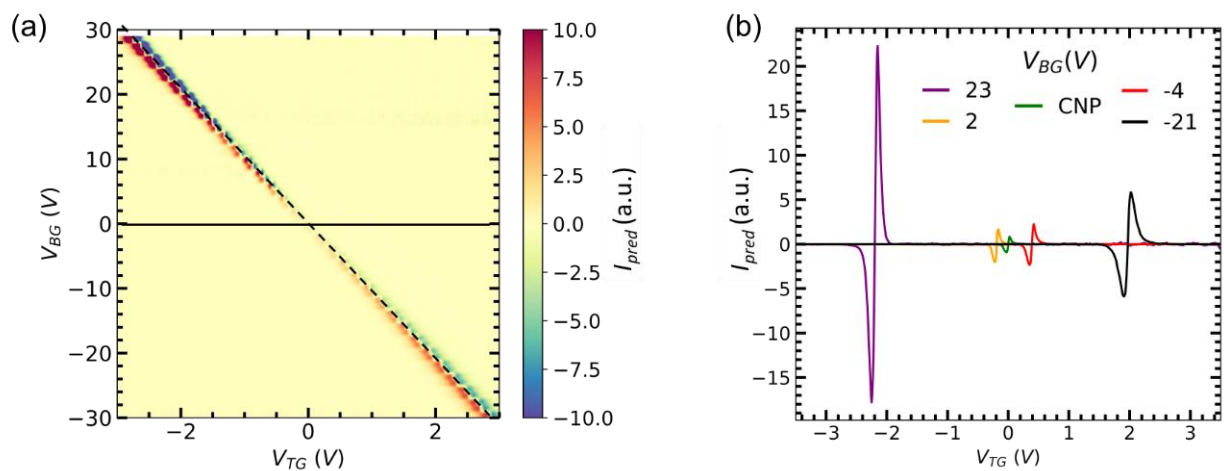

**Figure S3** (a) Predicted photocurrent ($I_{pred}$) mapping as a function of $V_{BG}$ and $V_{TG}$. The black horizontal (solid) and diagonal (dashed) lines correspond to the CNP of the non-top-gated and top-gated regions, respectively. (b) $I_{pred}$ as a function of top gate voltage $V_{TG}$ for five selected values of a back gate bias from panel (a). Expected values have been extracted from transport measurements at 10K following equation in the main text.

**Supplementary Material Note 4: Noise-Equivalent-Power.**

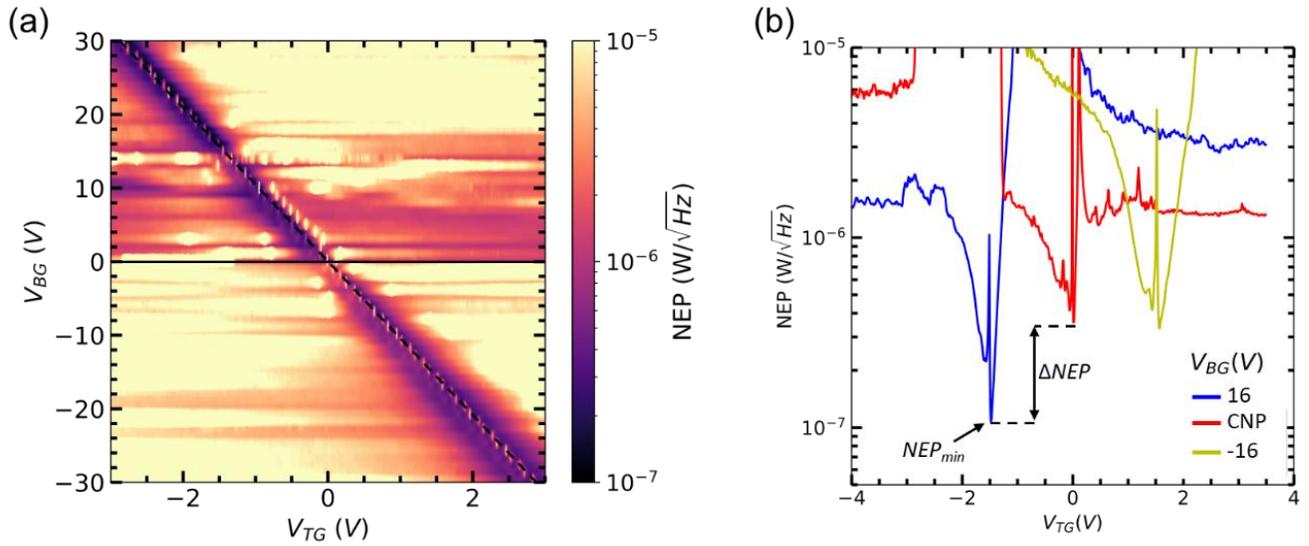

**Figure S4.** (a) Noise-Equivalent-Power ($NEP$) mapping as a function of $V_{TG}$ and $V_{BG}$ for an incident radiation of 0.3 THz at 10K. Horizontal (solid) and diagonal (dashed) lines correspond to the zero-charge-density states of the non-top-gated and top-gated regions, respectively. (b) $NEP$ as a function of top gate voltage $V_{TG}$ for three selected values of a back gate bias from panel (a). Minimizing the contribution of the access resistances yields to a reduction also in the minimum value of $NEP$.

## Supplementary Material Note 5: Room Temperature.

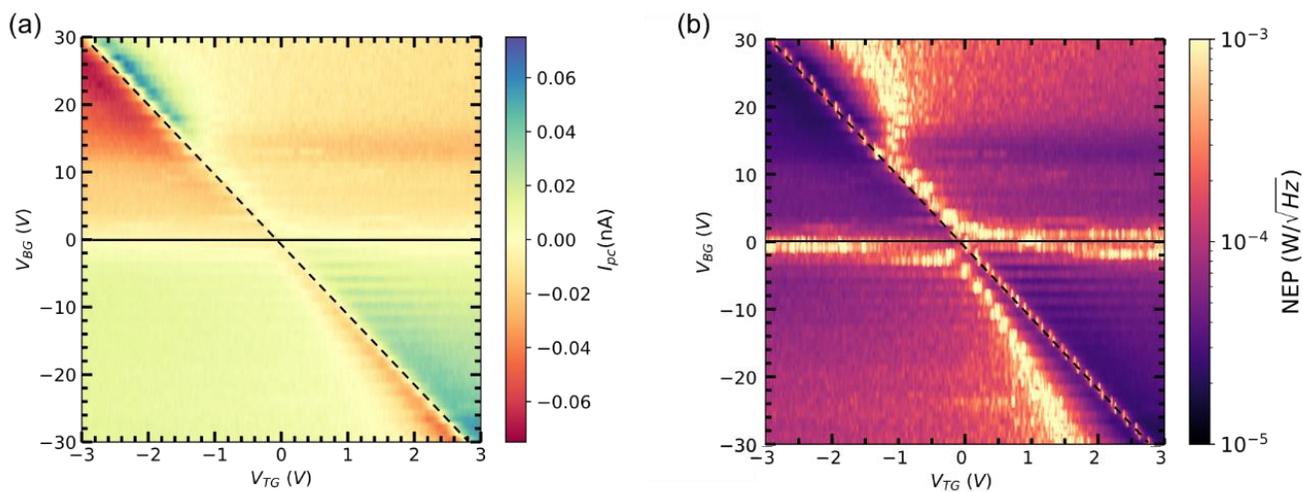

**Figure S5.** (a) Photocurrent mapping and (b) *NEP* mapping as a function of $V_{TG}$ and $V_{BG}$. Both panels have been obtained for an incident radiation of 0.3 THz at Room Temperature. Horizontal (solid) and diagonal (dashed) lines correspond to the zero-charge-density states of the non-top-gated and top-gated regions, respectively.

# REFERENCES

1.  Delgado-Notario JA, Clericò V, Diez E, Velázquez-Pérez JE, Taniguchi T, Watanabe K, Otsuji T, Meziani YM (2020) Asymmetric dual-grating gates graphene FET for detection of terahertz radiations. APL Photonics 5:066102. https://doi.org/10.1063/5.0007249

2.  Kim S, Nah J, Jo I, Shahrjerdi D, Colombo L, Yao Z, Tutuc E, Banerjee SK (2009) Realization of a high mobility dual-gated graphene field-effect transistor with Al2O3 dielectric. Appl Phys Lett 94:062107. https://doi.org/10.1063/1.3077021

3.  Gammelgaard L, Caridad JM, Cagliani A, MacKenzie DMA, Petersen DH, Booth TJ, Bøggild P (2014) Graphene transport properties upon exposure to PMMA processing and heat treatments. 2d Mater 1:035005. https://doi.org/10.1088/2053-1583/1/3/035005